\documentclass[conference]{IEEEtran}
\usepackage[left=0.625in, right=0.625in, bottom=0.9in, top=0.67in]{geometry}
\IEEEoverridecommandlockouts

\usepackage{amsfonts}
\usepackage{times}
\usepackage{graphicx}
\usepackage{epstopdf}
\usepackage{latexsym}
\usepackage{dsfont}
\usepackage{amssymb}
\usepackage{amsmath}
\usepackage{cite}
\usepackage{verbatim}
\usepackage{subfigure}

\newcommand{\figref}[1]{{Fig.}~\ref{#1}}

\def\bb0{{\mathbb{0}}}

\def\ba{{\mathbf{a}}}
\def\bb{{\mathbf{b}}}
\def\bc{{\mathbf{c}}}

\def\bff{{\mathbf{f}}}

\def\bh{{\mathbf{h}}}

\def\bt{{\mathbf{t}}}

\def\bv{{\mathbf{v}}}

\def\bx{{\mathbf{x}}}

\def\b0{{\mathbf{0}}}

\def\bH{{\mathbf{H}}}

\def\bQ{{\mathbf{Q}}}

\def\bV{{\mathbf{V}}}

\def\bX{{\mathbf{X}}}

\def\bbC{{\mathbb{C}}}

\def\bbE{{\mathbb{E}}}

\def\bbR{{\mathbb{R}}}

\def\bbZ{{\mathbb{Z}}}

\def\cN{\mathcal{N}}

\def\cQ{\mathcal{Q}}

\def\sf0{{\mathsf{0}}}

\usepackage{graphicx}
\usepackage{epstopdf}
\usepackage{enumerate}
\usepackage{algorithm}
\usepackage{amsmath}
\usepackage{mathrsfs}
\usepackage{float}
\usepackage{color}
\usepackage{makeidx}
\usepackage{bm}
\usepackage{cleveref}
\usepackage{url}
\usepackage{steinmetz}
\usepackage{varwidth}
\usepackage{soul}
\usepackage{bbm}

\newcommand{\sref}[1]{{Section}~\ref{#1}}

\def \rm {\mathrm}

\usepackage{amsfonts}
\usepackage{booktabs}
\usepackage{siunitx}
\usepackage{empheq}

\def\BibTeX{{\rm B\kern-.05em{\sc i\kern-.025em b}\kern-.08em
		T\kern-.1667em\lower.7ex\hbox{E}\kern-.125emX}}

\begin{document}
	\title{Camera Aided Reconfigurable Intelligent Surfaces: Computer Vision Based Fast Beam Selection}
	
	\author{Shuaifeng Jiang$^{\dagger}$, Ahmed Hindy$^{\ddagger}$, and Ahmed Alkhateeb$^{\dagger}$\\
		
		$^{\dagger}$ \textit{Arizona State University, Email: \{s.jiang, alkhateeb\}@asu.edu}\\
		$^{\ddagger}$ \textit{Motorola Mobility LLC (a Lenovo company), Email: ahmedhindy@motorola.com}
		\thanks{This work was supported by the National Science Foundation (NSF) under Grant No. 2048021.} }
	
	\maketitle
	\begin{abstract}
		Reconfigurable intelligent surfaces (RISs) have attracted increasing interest due to their ability to improve the coverage, reliability, and energy efficiency of millimeter wave (mmWave) communication systems. However, designing the RIS beamforming typically requires large channel estimation or beam training overhead, which degrades the efficiency of these systems. In this paper, we propose to equip the RIS surfaces with visual sensors (cameras) that obtain sensing information about the surroundings and user/basestation locations, guide the RIS beam selection, and reduce the beam training overhead. We develop a machine learning (ML) framework that leverages this visual sensing information to efficiently select the optimal RIS reflection beams that reflect the signals between the basestation and mobile users. To evaluate the developed approach, we build a high-fidelity synthetic dataset that comprises co-existing wireless and visual data. Based on this dataset, the results show that the proposed vision-aided machine learning solution can accurately predict the RIS beams and achieve near-optimal achievable rate while significantly reducing the beam training overhead.
	\end{abstract}

	\section{Introduction}
	Reconfigurable intelligent surfaces (RISs) are considered a low-cost and energy-efficient solution to enhance the performance and extend the coverage of future wireless communication systems \cite{Pan21, trichopoulos2022design}. RISs consist of a large number of reflecting elements that can modulate the amplitude and phase of the incident signals. When these elements are properly configured, the RIS can focus the incident signals toward the wireless receivers to compensate for the high pathloss and improve the signal-to-noise ratio (SNR). Moreover, these RISs can establish propagation paths that bypass the obstacles, which enables more reliable  connections. The highly-directional reflection beams can also reduce the inerference to other adjacent cells/users. Figuring out the right configuration of the RIS reflecting elements, however, requires sufficient channel knowledge. Acquiring this knowledge is very challenging for the large-dimensional RISs and requires high channel estimation and beam training overhead. \cite{taha2021enabling}.
		\begin{figure}[t]
		\centering
		\includegraphics[width=1 \linewidth]{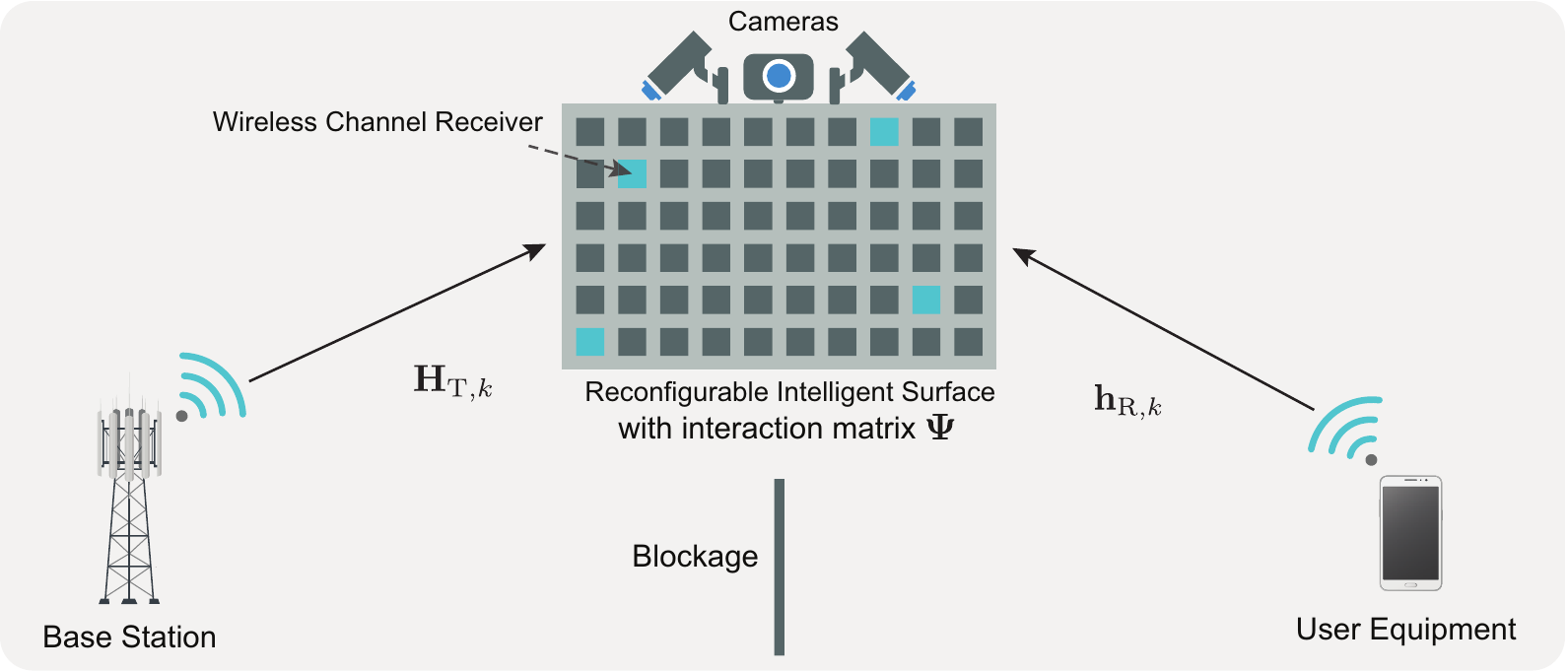}
		\caption{The system model consists of the BS, the UE, and the RIS. The link between the BS and the UE is blocked, and their communication is aided by the RIS. The RIS is equipped with cameras to guide its beam selection.}
		\label{fig:system model}
	\end{figure}

	Prior work has investigated the RIS beamforming design and attempted to reduce the channel estimation and beam training overhead. For example,  \cite{yang2020intelligent} proposed to divide the reflecting elements into groups and estimate the effective channel for each group. \cite{jensen2020optimal} developed a channel estimation approach that relies on varying the RIS phase shift configurations with each pilot symbol. Although the estimated channels can be used to design promising RIS beamforming performance, it leads to large channel estimation complexity and overhead that scales with the number of RIS reflecting elements and the BS/user antennas. Alternatively, the RIS can configure its elements by beam training using a pre-defined codebook \cite{trichopoulos2022design}. Prior work has studied reducing the beam training overhead by ways such as hierarchical beam search \cite{ning2021terahertz} and RIS sub-array clustering \cite{you2020fast}. The beam sweeping time of these solutions, however, still scales with the number of reflecting elements and leads to high latency, making it hard for these systems to support mobile scenarios. In another context, integrating	sensing capabilities at the infrastructure and mobile users to aid the wireless communication decisions has been recently attracting interest for different use cases \cite{jiang2021computer,charan2021vision,tx_id,Demirhan_mgazine_radar}. In particular, prior work studied using visual sensing information to improve beam tracking \cite{jiang2021computer}, blockage prediction \cite{charan2021vision}, and codebook design \cite{tx_id}. Visual sensors provide fine-grained spatial information for LoS objects, which is particularly suitable for mmWave RISs as they often aid the communications via LoS paths.

	In this paper, we propose a novel RIS beam selection approach leveraging the visual sensory data collected by cameras deployed at the RIS. The proposed vision-aided beam selection approach can significantly reduce the beam training overhead and achieve near-optimal achievable rate. The contribution of the paper is threefold. (i) We propose to equip the RIS with visual sensing capability to guide the RIS beam selection process, and mathematically formulate the vision-aided RIS beam reflection problem. (ii) We develop a machine learning framework for the RIS beam selection task. The ML framework first detects candidate users using computer-vision object detectors and then exploits a designed neural network to predict the RIS candidate beam set. (iii) We develop a neural network architecture that improves the RIS beam prediction performance and features good learning capabilities. To evaluate the performance of the developed solution, we build a high-fidelity synthetic dataset comprising wireless and visual data for a realistic RIS-aided wireless communication scenario. The evaluation results highlight the potential of the proposed vision-aided RIS beam selection approach in reducing the beam training overhead and enabling the RIS operation in highly-mobile scenarios.

	\section{System and Channel Models}\label{System and Channel Models}
	Here, we describe the adopted system and channel models.

	\subsection{System Model}
	We consider a communication system where an RIS is deployed to aid the communication between a BS and a UE. For ease of exposition, we assume that a blockage exists between the BS and the UE. Consequently, the BS can only communicate with the UE through the RIS. It is important to mention here, however, that the proposed beam selection in this paper can generally also apply to scenarios with direct links between the BS and the UE. We assume that the BS has an antenna array of $N$ elements, and the UE is equipped with a single antenna. The RIS has $M$ passive reflecting elements that can shift the phase of the incident signals. Further, the RIS is assumed to be equipped with RGB cameras to obtain sensing information about the surrounding environment. There are three differently-oriented cameras deployed at the center of the RIS surface. These cameras provide a central view and two side views of the surrounding environment. Here, we assume that the area covered by the RGB cameras is within the coverage area of the BS.

	For the downlink communication, we adopt orthogonal frequency-division multiplexing (OFDM) with $K$ subcarriers. Let $\bH_{{\rm T}, k}\in \bbC^{M \times N}$ and $\bh_{{\rm R}, k}\in \bbC^{M \times 1}$ denote the channel matrix from the BS to the RIS and the channel vector from the UE to the RIS at the $k$-th subcarrier, respectively. If the BS transmits a downlink signal $s_k \in \bbC$ on the $k$-th subcarrier, then, we can write the received signal at the UE as
	\begin{align}\label{eq:signal model}
		y_k &= \bh_{{\rm R}, k}^{T} {\bf \Psi} \bH_{{\rm T}, k} \bff s_k + n_k,
	\end{align}
	where $\bff \in \bbC^{N\times 1}$ denotes the beamforming vector of the BS. The transmitted signal $s_k$ satisfies the power constraint $\bbE\left[|s_k|^2\right] = \frac{p_t}{K}$ with $p_t$ representing the average total transmit power of the BS. $n_k \sim \cN_{\bbC}(0, \sigma^2_n)$ is the receive additive white Gaussian noise at the UE. ${\bf \Psi} \in \bbC^{M\times M}$ denotes the RIS interaction matrix which captures the adopted RIS reflection configuration. The matrix ${\bf \Psi}$ is diagonal and can be written as
	\begin{align}
		{\bf \Psi} = \text{diag}({\bm \psi})=\text{diag}(\psi_1, \hdots, \psi_M),
	\end{align}
	where ${\bm \psi}=\left[\psi_1, \hdots, \psi_M\right]^T$ is the diagonal vector of $\bf \Psi$ with $\psi_m \in \bbC$ representing the $m$-th reflecting element of the RIS. We call $\bm \psi$ the reflecting beamforming vector of the RIS. To account for the practical constraint of quantized phase shifters \cite{alkhateeb2014channel}, we assume that the RIS adopts a pre-defined reflecting beam codebook $\cQ = \{\cQ_1, \hdots, \cQ_{|\cQ|}\}$, \textit{i.e.}, ${\bm\psi} \in \cQ$. Note that, the RIS beam $\bm\psi$ satisfies $|\psi_m|^2=1$ to capture the constant modulus phase-only constraint. Note that the same reflection vector ${\bm \psi}$ is applied to all the K subcarriers due to the time-domain implementation.
	\subsection{Channel Model}
	We adopt a wideband block-fading geometric channel model for the channels $\bH_{\mathrm{T},k}$ and $\bh_{\mathrm{R},k}$. With this model, if $\bh_{\mathrm{R},k}$ consists of $L$ clusters, and each cluster $\ell\in[1,L]$ contributes with one ray of time delay $\tau_\ell \in \bbR$, then the delay-d channel vector between the UE and the RIS can be written as
	\begin{equation}\label{eq:delay-d}
		\bh_{\mathrm{R},d} = \sqrt{\frac{M}{\rho}} \sum_{\ell=1}^L \alpha_\ell p(dT_s - \tau_\ell) \ba(\phi^R_{\ell}, \theta^R_{\ell}),
	\end{equation}
	where $\rho$ denotes the pathloss and $p(\tau)$ denotes the pulse shaping function which represents a $T_S$-spaced signaling evaluated at $\tau$ seconds, $\ba(\phi^R_{\ell}, \theta^R_{\ell})$ is the array response vector of the RIS. $\phi^R_{\ell}$ and $\theta^R_{\ell}$ are the corresponding azimuth and elevation angles of arrival (AoA) associated with the $\ell$-th cluster. $\alpha_\ell \in \bbC$ is the complex gain of the $\ell$-th cluster.

	Given the delay-d channel in \eqref{eq:delay-d}, the frequency domain channel vector at subcarrier $k$ can be written as
	\begin{equation} \label{eq-channel}
		\bh_{\mathrm{R},k} = \sum_{d=0}^{D-1} \bh_{\mathrm{R},d} e^{-j\frac{2\pi k}{K} d},
	\end{equation}
	where $D$ is the maximum delay of the channel corresponding to the delay spread. The channel  $\bH_{\mathrm{T},k}$ is similarly defined.
	\section{Vision-Aided RIS Beam Selection Problem Formulation}\label{Vision-Aided RIS Operation}
	In this paper, we propose to enable RIS beam selection and reduce beam training overhead by exploiting the visual sensors. To start, we adopt the achievable rate as the performance metric of interest. Given the system and signal models in Section \ref{System and Channel Models}, the downlink achievable rate for the adopted RIS-based communication system can be written as
	\begin{align}
		R &= \frac{1}{K}\sum_{k=1}^{K}\log_2\left( 1 + \textrm{SNR}\left| \bh^T_{{\rm R}, k} {\bf \Psi} \bH_{{\rm T}, k} \bff \right|^2 \right)\nonumber\\
		&= \frac{1}{K}\sum_{k=1}^{K}\log_2\left( 1 + \textrm{SNR}\left| \left(\bh_{{\rm R}, k}\odot \bH_{{\rm T}, k} \bff \right)^T{\bm \psi}\right|^2 \right),
	\end{align}
	where SNR$=\frac{p_t}{K\sigma_n^2}$ denotes the signal-to-noise ratio. The $\odot$ denotes the dog product of two vectors.
	Therefore, for a given BS beamforming vector $\bff$, the optimal RIS reflecting beam for this pair of BS and UE is the one that maximizes the achievable rate. Since the RIS exploits a pre-defined reflecting beam codebook, the optimal RIS beam is uniquely indicated by its beam index $q\in\{1,\hdots,|\cQ|\}$ in the pre-defined codebook.  The optimal RIS beam index that maximizes the achievable rate is given by
	\begin{align}\label{eq:best_beam}
		{q}^\star = \underset{q\in\{1,\hdots,|\cQ|\}}{\arg\max} \, R.
	\end{align}
	Then, our objective is to find the optimal RIS beam index $q^\star\in\{1,\hdots,|\cQ|\}$ in \eqref{eq:best_beam} with the smallest number of trials.
	\begin{figure*}[]
		\centering
		\includegraphics[width=0.95\linewidth]{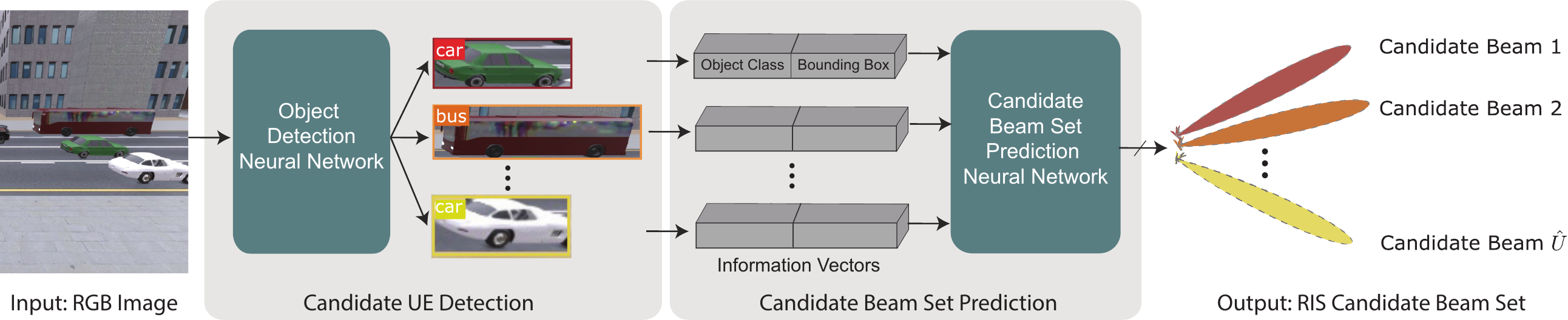}
		\caption{Overview of the proposed ML framework for the RIS candidate beam set prediction. The RIS uses an object detector to detect the candidate UEs from the visual information. Then the information of the candidate UEs is used to predict the candidate beams.}
		\label{fig:framework}
	\end{figure*}

	Conventional approaches to obtain optimal RIS beams may require high pilot or beam training overhead, which degrades the system efficiency. To tackle this problem, we propose to utilize visual sensors (cameras) at the RIS to aid the beam selection. Let $\bX \in \bbR^{w\times h \times 3}$ denote one RGB image captured by the camera at the RIS with $w$ and $h$ denoting the width and height of $\bX$. Since one image can contain multiple (candidate) UEs, the objective of the vision-aided RIS beam selection is then to accurately predict a set of candidate RIS beams that correspond to the candidate UEs in the image. More specifically, for each candidate UE, we aim to predict the RIS beam $q^\star$ that satisfies \eqref{eq:best_beam}. The optimal RIS candidate beam set of the image $\bX$ can then be defined as
	\begin{align}\label{eq:best_beamset}
		{\bQ}^\star_\bX =\left\{{q}_1^\star, \hdots, {q}_U^\star \right\},
	\end{align}
	where $U$ is the total number of ground-truth candidate UEs in the image, and $q_u^\star\ (u=1,\hdots, U)$ is the optimal RIS beam index for the $u$-th ground-truth UE.
	
	The objective is then to find a function that can map the image $\bX$ captured at the RIS to the optimal RIS candidate beam set. The exact mathematical relation between $\bX$ and $\bQ^\star_\bx$, however, is very difficult to characterize because it depends on the channel model, the visual model, and the layout of the environment around the BS, the UE, and the RIS. This motivated utilizing ML to approach the mathematical relation between $\bX$ and $\bQ^\star_\bx$ in a data-driven manner. The optimal ML model $f^\star(\cdot;\boldsymbol{\Theta}^\star)$ can be expressed as
	\begin{align}\label{eq:optim_ml}
		f^\star(\bX;\boldsymbol{\Theta}^\star) ={\bQ}^\star_\bX,
	\end{align}
	where $\boldsymbol{\Theta}^\star$ denotes the parameters of the optimal ML model. Next, we elaborate on the adopted ML model in Section~\ref{Deep Learning Modeling}.

	\section{Deep Learning Modeling}\label{Deep Learning Modeling}
	In this section, we elaborate on the proposed ML framework for the vision-aided RIS beam selection task. As shown by \figref{fig:framework}, the ML framework first detects the candidate UEs based on the visual sensing information using a computer-vision object detector. Then the ML framework employs a NN to predict the RIS candidate beam set given the information of the detected candidate UEs.
	\subsection{Candidate UE detection}\label{Candidate UE detection}
	The first component of the ML framework is a computer-vision object detector that is used to detect the candidate UEs. Since the wireless environment is typically changing quickly, the object detector in the proposed framework for RIS candidate beam set prediction needs to satisfy an essential requirement: the capability to produce fast object detections of high quality. To that end, we select the well-known YOLOv3 object detector \cite{yolov3} because of its fast prediction speed and high accuracy. To obtain a YOLOv3 model that can accurately detect candidate UEs with the visual information at the RIS, there is no need to train the model from scratch. Instead, we start from a pre-trained model and fine-tune it on the images from the RIS's cameras. Given an input image, the YOLOv3 model outputs a class index $c\in \bbZ$, and a bounding box vector $\bb\in \bbR^{4 \times 1}$ for each detected object. The bounding box vector consists of the x-center, the y-center, the width, and the height of the bounding box.
	\subsection{Candidate Beam Set Prediction} \label{Candidate Beam Set Prediction}
	The YOLOv3 object detector provides the class and bounding box information for each detected candidate UEs in an image. Based on this information, we now design a NN that can predict the RIS candidate beam set. Next, we will describe the key components of the proposed NN for the RIS candidate beam set prediction, namely the input/output representation, the NN architecture, and the loss function and learning model.

	\noindent \textbf{Input/Output Representation and Normalization:}
	Given an image, the YOLOv3 model detects $\hat{U}$ candidate UEs. For each candidate UE, the YOLOv3 model outputs a class index $c\in \bbZ$, and a bounding box $\bb\in \bbR^{4 \times 1}$. To make the training process of the NN faster and more stable,
	we convert the class $c$ to a one-hot vector $\bar{\bc}$, and we normalize the bounding box $\bb$ by the size of the image, $w$ and $h$. The normalized bounding box is denoted by $\bar{\bb}$. Then the one-hot representation of the class $\bar{\bc}$ and the normalized bounding box $\bar{\bb}$ are concatenated to form the candidate UE information vector $\bv = [{\bar{\bc}}^T, \bar{\bb}^T]^T$. Finally, the input matrix $\bV$ to the NN architecture is written as
	\begin{align}
		\bV = [\bv_1, \hdots, \bv_{\hat{U}}, {\bf 0}, \hdots, {\bf 0}].
	\end{align}
	Note that we pad $(U_{max}-\hat{U})$ zero-vectors since the number of detected UEs varies from image to image. $U_{max}$ denotes the maximum number of candidate UEs that exist in any image.
	\par
	To construct the desired output of the NN, we first obtain the optimal RIS beam set $\bQ^\star_\bX$ corresponds to the image $\bX$ as shown by \eqref{eq:best_beamset} with exhaustive search over the codebook $\cQ$. Then we convert $\bQ^\star_\bX$ into a multi-hot vector $\bt^\star=\left[t_1^\star,\hdots, t^\star_{|\cQ|}\right]$. For any element of $\bt^\star$, $t_q^\star$ satisfies
	\begin{align}\label{thresh}
		t_q^\star=
		\begin{cases}
			1 & q \in \bQ^\star_\bX \\
			0 & \text{otherwise},
		\end{cases}
	\end{align}
	where $q$ is a RIS beam index in the RIS beam codebook $\cQ$.

\begin{figure}[t]
		\centering
		\includegraphics[width=1\linewidth]{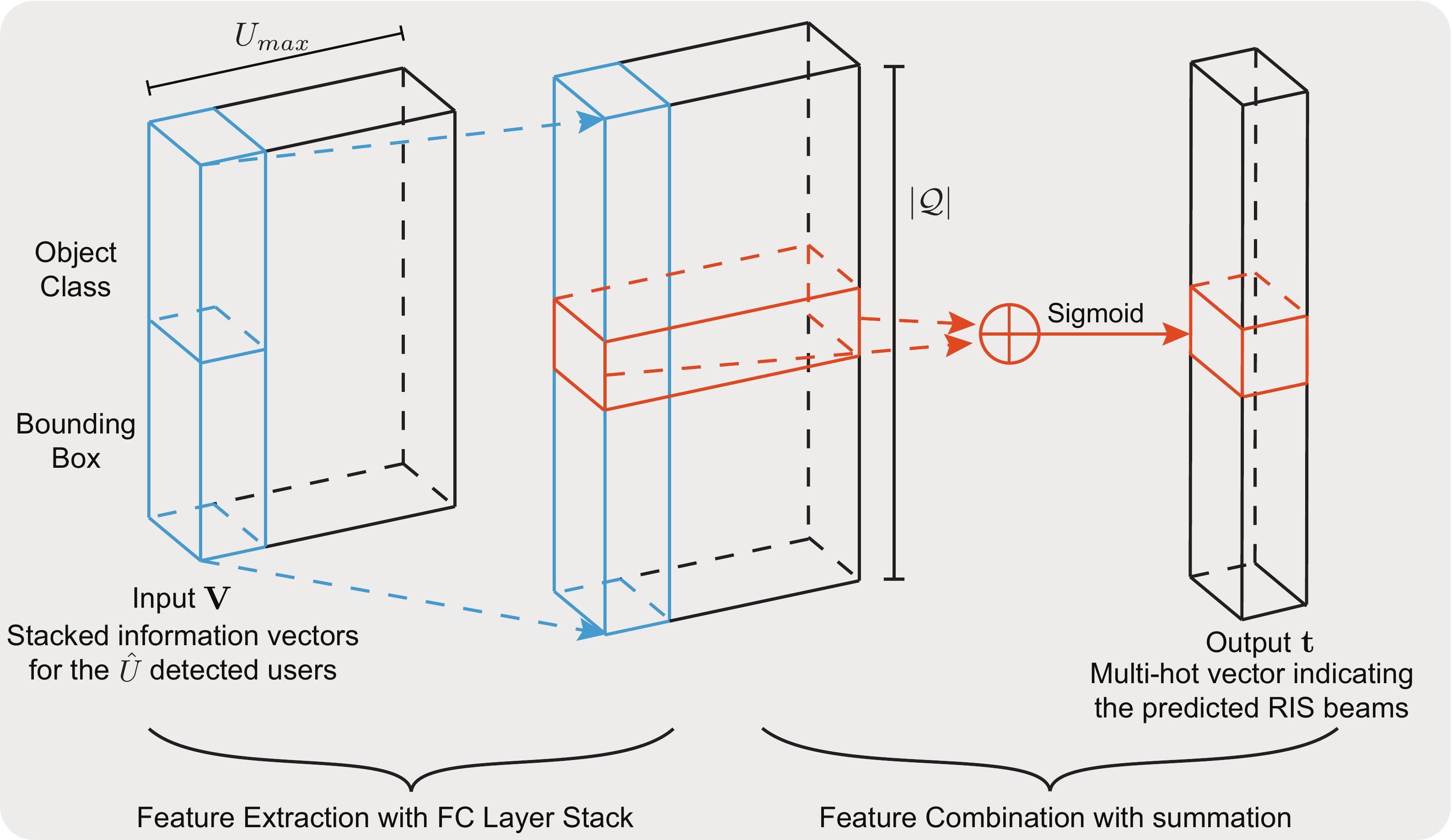}
		\caption{This figure shows the proposed NN architecture for predicting candidate beam set from detected UE information.}
		\label{fig:NN_arch}
	\end{figure}

	\noindent \textbf{NN Architecture:}
	As shown by Fig. \ref{fig:NN_arch}, we propose an NN architecture that predicts the RIS candidate beams from the class and bounding box information of all detected candidate UEs. The proposed NN first applies the same stack of fully connected NN layers on each candidate UE's information vector (each column of $\bV$) to extract high-level features. These fully-connected layers adopt the ReLU activation function. After this feature extraction, each input candidate UE information vector is transformed to a $|\cQ|$-dimensional vector. Then these $|\cQ|$-dimensional vectors are combined by a summation operation. Since the desired output of the NN is a multi-hot vector, the sigmoid activation function is applied to the combined vector to restrict its value into the range $(0,1)$. Lastly, we obtain the output vector of the NN, $\bt \in \bbR^{|\cQ|\times 1}$.

	The proposed NN architecture has two advantages for predicting the RIS candidate beam set from the candidate UE information vectors.
	First, it reuses the same stack of fully connected layers to extract features from different candidate UE information vectors. This aligns with the intuition that all candidate UEs are equivalent for the NN architecture, thus, they should be processed in the exact same way. Reusing this same stack of fully connected layers also reduces the complexity of the proposed NN architecture, which stabilizes the training process and reduces the computational complexity of the inference process.
	Second, the output of the proposed NN architecture does not rely on the order of the input candidate UE information vectors. This is achieved by reusing the same stack of fully connected layers and the summation operation that combines features from all candidate UE information vectors. Therefore, the proposed NN architecture can be more robust by not overfitting to the order of the input candidate UE information vectors.
	We validate our intuitions on the NN architecture by numerical results presented in \sref{sim result1}.

	\noindent \textbf{Loss Function and Learning Model:}
	The NN is designed to predict the RIS candidate beam set within a beam codebook for all candidate UEs. This can be modeled as a multi-class classification problem. Therefore, we adopt a classification learning model. We train the NN by supervised learning and employ the cross-entropy loss function.
	\section{Dataset and Performance Metrics}\label{setup}
	In this section, we first explain the considered simulation setup. Second, we elaborate on the generated dataset that is later used to train and evaluate the NN. Then, we introduce the performance metrics for the RIS beam set prediction.
	\subsection{Simulation Setup}
\begin{figure}[t]
		\centering
		\includegraphics[width=0.95\linewidth]{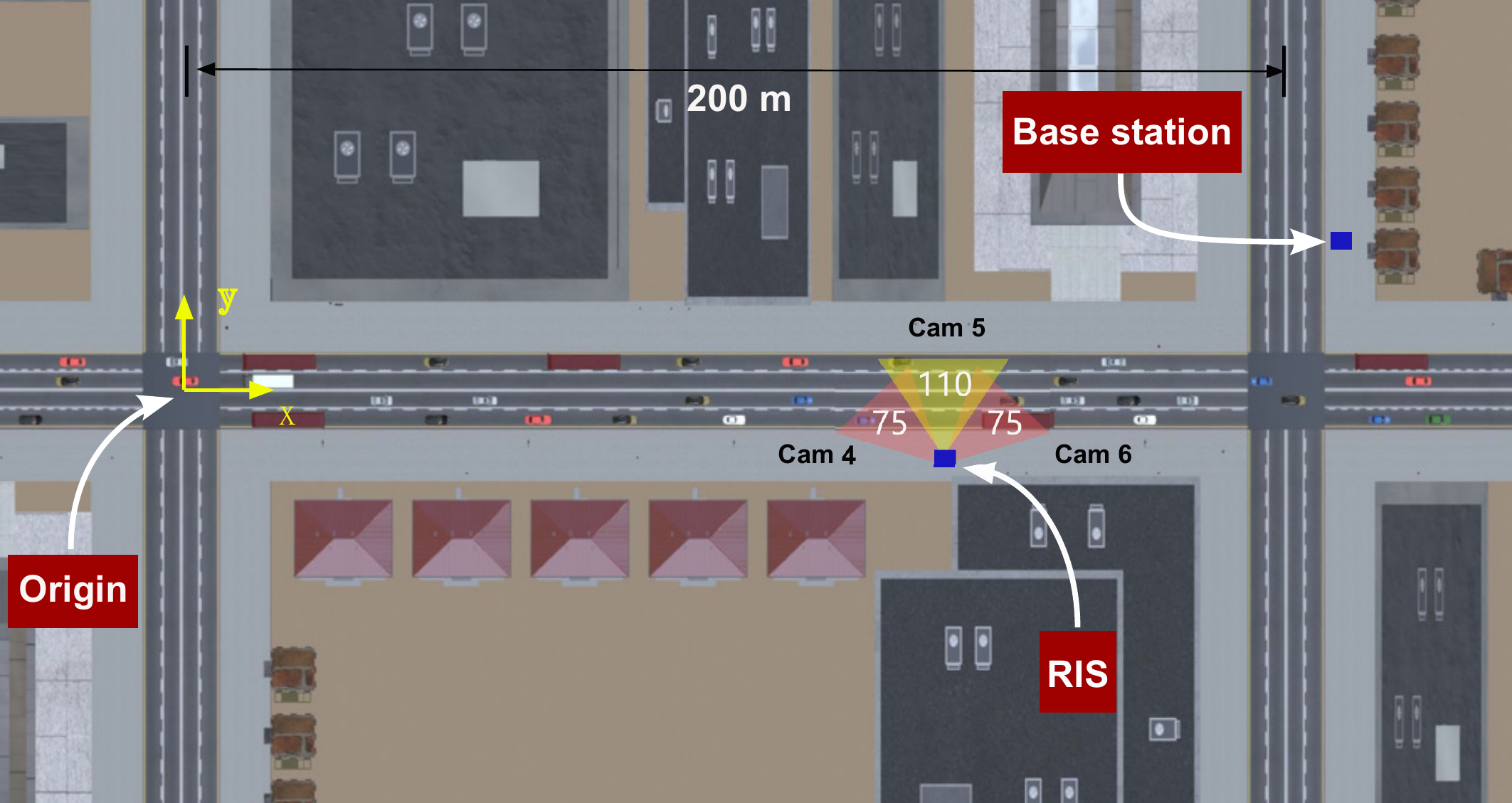}
		\caption{The top-view of the simulation scenario depicting a downtown area.}
		\label{fig:viwi}
	\end{figure}

	We investigate utilizing visual sensing-based perception to enable RIS beam selection. Hence, realistic wireless and visual modelings are essential for our simulation. To that end, we generate the training and test data with the ViWi dataset \cite{viwi}. The ViWi dataset provides co-existing wireless and visual data based on accurate ray tracing. It comprises sequences of RGB frames, beam indices, and user link statuses. They are generated from a large simulation of a synthetic outdoor environment depicting a downtown street with multiple moving objects. To simulate our RIS-aided communication system, we construct a new scenario based on the ViWi scenario 1. A top view of our scenario is presented in Fig. \ref{fig:viwi}. The BS is located on the vertical street at the upper right. The UEs are the moving vehicles on the main street. The RIS is installed at the side of the main road to aid communications between the BS and the UEs. In the simulation, for simplicity, we assume the BS and the UEs to be single-antenna. The RIS is equipped with a uniform planar array (UPA) with 32 columns and 8 rows of reflecting elements, \textit{i.e.}, the total number of reflecting elements of the RIS is $256$. Three cameras (``Camera 4", ``Camera 5" and ``Camera 6") are deployed at the RIS as shown in Fig. \ref{fig:viwi}. The central ``Camera 5" has a $110^\circ$ field of view while the side cameras' field of views are $75^\circ$. In the following simulations, we focus on the UEs in the views of ``Camera 4" and ``Camera~5".
	\subsection{Data Generation}
	We first generate 10000 scenes with the ViWi dataset. Then, for each camera, one scene consists of an image $\bX$, and the channels of all UEs in the view of the camera. Note that we only keep the data where the UEs do not have LoS paths with the BS. We obtain the optimal RIS candidate beam set ${\bQ}^\star$ for each image according to \eqref{eq:best_beam} and \eqref{eq:best_beamset} by exhaustively searching over the codebook $\cQ$. The optimal RIS candidate beam set for each image is then converted to the multi-hot representation $\bt^\star$. Then $\bX$ and $\bt^\star$ form a data point $(\bX, \bt^\star)$. Lastly, we obtain two datasets for ``Camera 4" and ``Camera 5", respectively. As discussed in \ref{Candidate UE detection}, we fine-tune the YOLOv3 model on 500 images randomly selected from each camera's data. The bounding boxes and object classes for all visible vehicles of all images are manually labeled.	After fine-tuning the YOLOv3 model, we generate the datasets to train the beam set prediction NN. First, we apply the fine-tuned YOLOv3 to obtain the candidate UE information vectors of all images. After that, each data point $\left(\bX, \bt^\star\right)$ is converted to $\left(\bV, \bt^\star\right)$. The two datasets for ``Camera 4" and ``Camera 5" are split into training and test sets using an $80\%$-$20\%$ data split. Note that two NNs are separately trained and evaluated on the ``Camera 4" and ``Camera~5" datasets since they have different camera angles.
		\begin{figure}[t]
		\centering
		\includegraphics[width=1\linewidth]{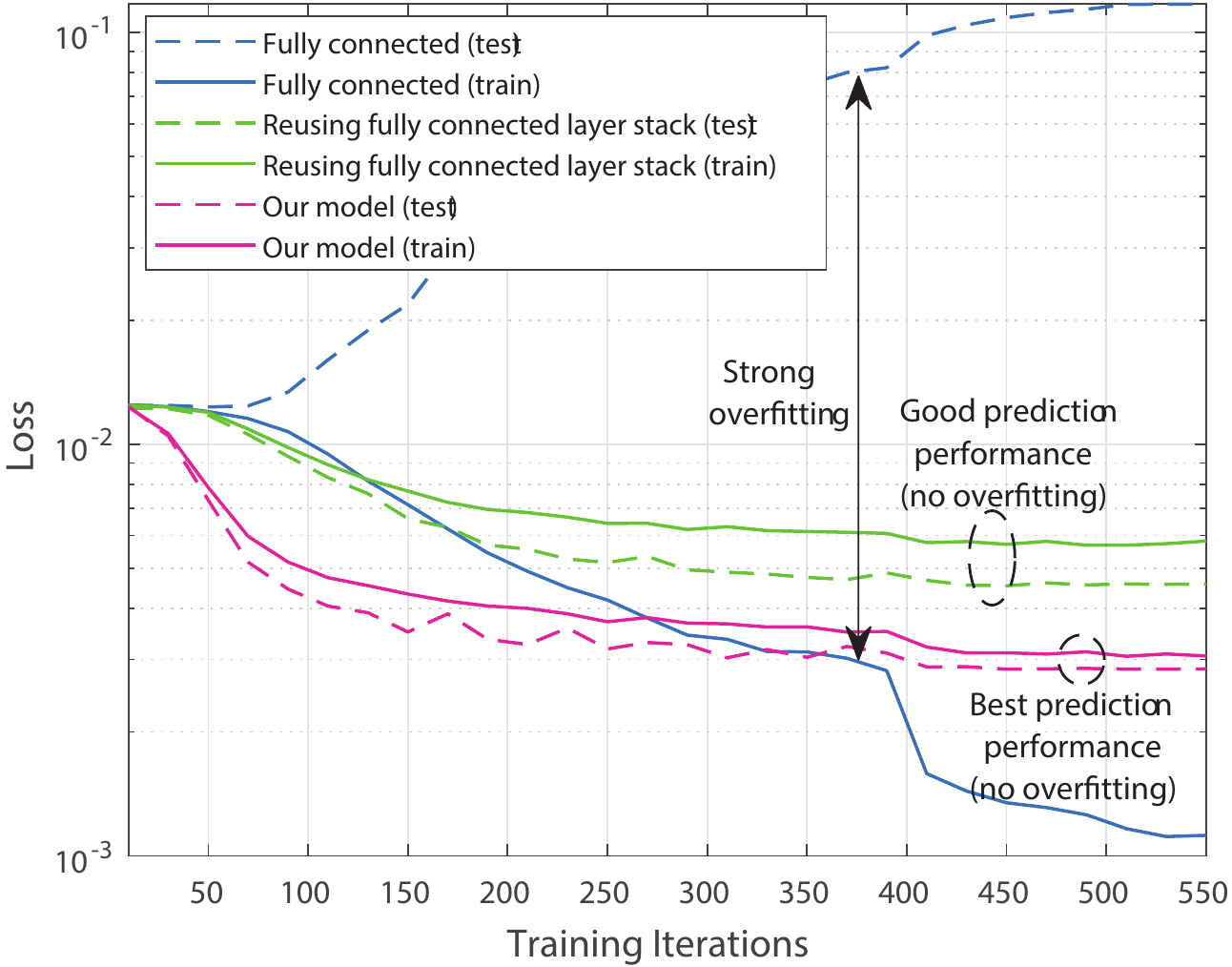}
		\caption{This figure compares the learning curves (train/test) of the proposed NN structure and two baseline models trained on the ``Camera 5" dataset.}
		\label{fig:learning_curve}
	\end{figure}

	\subsection{Performance Metrics} \label{subsec:metrics}
	Here, we present the metrics followed to evaluate the quality of the NNs' RIS beam set prediction. In the NNs' output vector $\bt$, each element represents a promising score of the corresponding beam in $\cQ$. To evaluate the prediction performance, we first apply the following unit step function on each element of the output vector, $f_{\mathrm{step}}(x)= u(x-\delta)$,	where we use $\delta=0.5$ as the threshold. By applying this threshold to $\bt$, we obtain $\hat{\bt}$. The predicted RIS candidate beam set $\hat{\bQ}$ can then be written as
	\begin{align}\label{eq:choose_beam}
		\hat{\bQ} = \left\{ q| \hat{t}_q =1\right\},
	\end{align}
	where $\hat{t}_q$ is the $q$-th element of $\hat{\bt}$. The metrics adopted to evaluate the performance of the RIS candidate beam set prediction are the accuracy and the recall as given by
	\begin{subequations}
		\begin{align}
			\mathrm{Acc} &= \frac{1}{N_{test}}\sum_{i=1}^{N_{test}} \frac{\left|\bQ^\star_i \cap \hat{\bQ}_i\right|}{\left|\hat{\bQ}_i\right|} \\
			\mathrm{Recall} &= \frac{1}{N_{test}}\sum_{i=1}^{N_{test}} \frac{\left|\bQ^\star_i \cap \hat{\bQ}_i\right|}{\left|\bQ^\star_i \right|},
		\end{align}
	\end{subequations}
	where $N_{test}$ is the number of data samples in the test dataset. $\bQ^\star_i$ and $\hat{\bQ}_i$ denote the optimal and the predicted RIS candidate beam set for the $i$-th data sample, respectively.

	\begin{figure}[t]
		\centering
		\includegraphics[width=.96\linewidth]{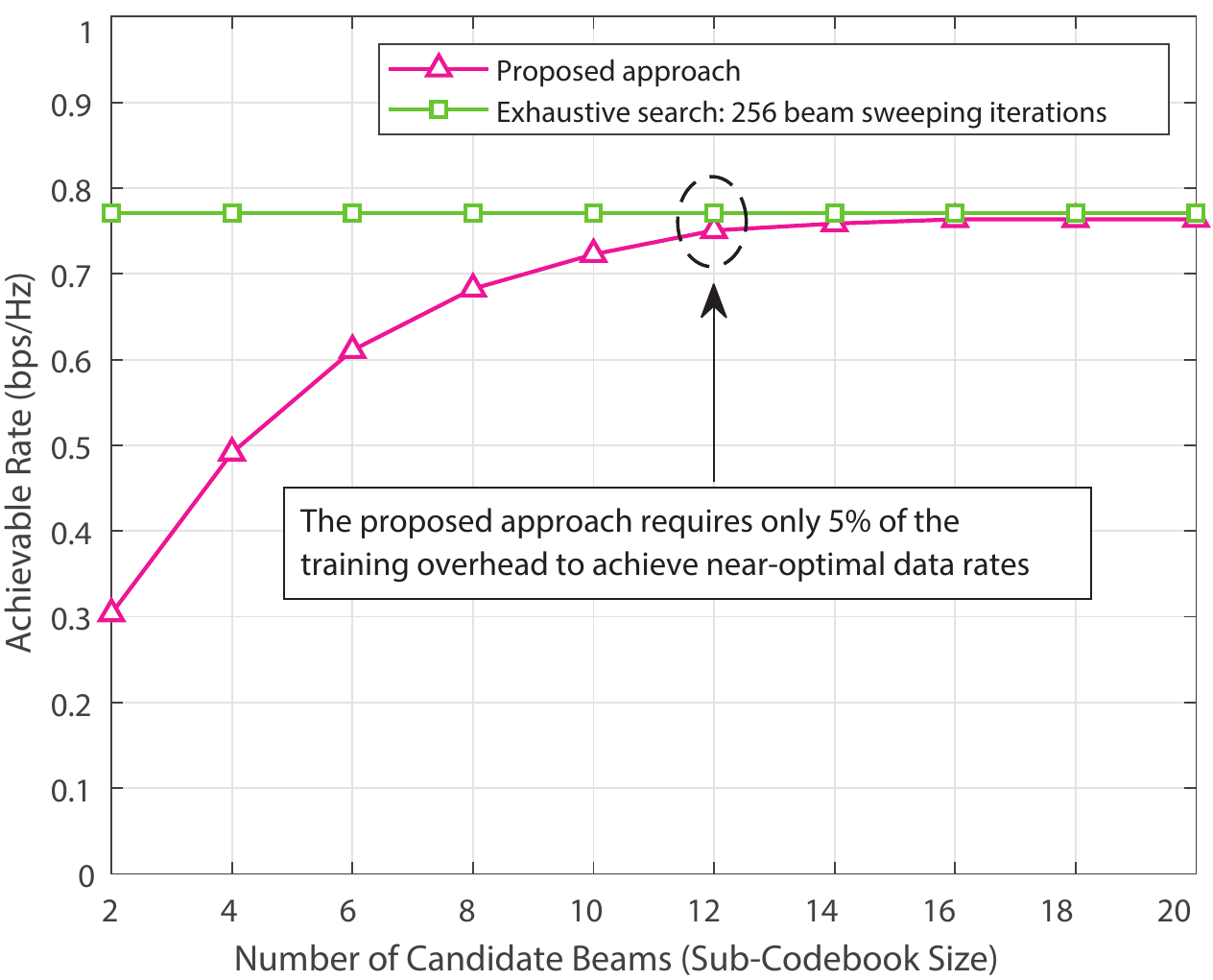}
		\caption{The achievable rate of the proposed vision-aided RIS beam selection with increasing size of the RIS candidate beam set. The upper bound achievable rate is obtained by the exhaustive search over $256$ RIS beams.}
		\label{fig:ach_rate_beam_set_size}
	\end{figure}

	\section{Simulation Results}\label{Simulation}
	In this section, we evaluate the performance of the proposed vision-aided RIS beam selection.
	\subsection{Does the Proposed Neural Network Learn Better?}\label{sim result1}
	In \sref{Candidate Beam Set Prediction}, we mentioned two key features of the proposed NN structure: (i) Reusing the fully connected layer stack on all the UE information vectors, and (ii) combining the information of the detected UEs by the summation operation. These two features are expected to improve the performance of the candidate beam set prediction. To analyze the effectiveness of the proposed NN structure and verify the intuitions used in its design, we study the training process of the NN. Fig.~\ref{fig:learning_curve} presents the learning curves of the proposed NN structure compared with two variants trained on the \mbox{``Camera 5"} dataset. The first variant adopts vanilla fully connected NN. The second variant reuses the same fully connected layer stack on the information vectors from all candidate UEs, but it concatenates the resulting high-level feature vectors instead of applying the summation operation. From the training and the test loss in Fig. \ref{fig:learning_curve}, we see that the vanilla fully connected NN overfits to the training dataset and its loss diverges on the test dataset. For the second variant, the test loss can converge along with training iterations. This indicates that \textbf{reusing the fully connected layer stack stabilizes the training process}. The proposed NN structure achieves the lowest loss on the test dataset. This implies that \textbf{combining the information from different UEs with the summation operation improves the generalization performance of the model}. These results highlight the effectiveness of the proposed NN architecture.

	\subsection{Can the Proposed NN improve candidate beam prediction?}\label{sim result2}
	In Table \ref{tb:finall_performance0}, we present the accuracy and recall performance of the proposed NN structure compared with its two variants. Our proposed NN structure achieves $94.2\%$ and $92.9\%$ on the ``Camera 4'' dataset for accuracy and recall, respectively. On the ``Camera 5'' dataset, the accuracy and recall performance of the proposed NN structure are $92.1\%$ and $86.4\%$. These results highlight that the proposed NN structure can accurately predict the RIS candidate beam set from the UE information extracted by the YOLOv3 model. Comparing the proposed NN structure with other two variants, it can be seen that reusing the fully connected stack offers significant improvements, in terms of accuracy and recall, on both datasets. \textbf{For the dataset of ``Camera 5'', the accuracy increases by $\bf{59.5\%}$. On top of that, by reusing the fully connected stack and combining information of candidate UEs with the summation operation, our NN structure results in the highest performance on the test datasets. The recall performance for the ``Camera 5' dataset is improved by $\bf{15.1\%}$.} This again emphasizes that the two key features of the proposed NN structure help stabilize the training process and achieve better performance in the beam set prediction.
	\begin{table}[t]
		\normalsize
		\setlength\tabcolsep{5pt}
		\caption{\label{tb:finall_performance0}Accuracy and recall performance of three different NNs.}
		\centering
		\footnotesize
		\begin{tabular}{lllll}
			\toprule
			\multicolumn{1}{c}{Model}& \multicolumn{2}{c}{Accuracy} & \multicolumn{2}{c}{Recall}\\
			\cmidrule(lr){1-1} \cmidrule(r){2-3} \cmidrule(lr){4-5}
			& {Cam4} & {Cam5} & {Cam4} & {Cam5}\\
			Fully connected& {71.2\%} & {32.6\%} & {70.1\%} & {25.5\%} \\
			Reuse FC stack&{92.3\%} & {92.1\%} & {87.6\%} &{71.3\%}\\
			Reuse FC stack \& sum operation& {\bf 94.2\%} &{\bf 92.9\%} &{\bf 92.1\%} & {\bf 86.4\%} \\
			\bottomrule
		\end{tabular}
	\end{table}
	\subsection{How Much Beam Training Overhead is Required?}
	Here, we evaluate how much training overhead is required using the proposed vision-aided RIS beam selection. We assume that the RIS is controlled by the BS to perform RIS beam training with a sub-codebook of $\cQ$ given the predicted RIS candidate beam set. The RIS sub-codebook is constructed by the $k$ beams corresponding to the top-$k$ highest value in $\bf t$, the output vector of the ML framework. In \figref{fig:ach_rate_beam_set_size}, we investigate the effect of the size of the RIS sub-codebook on the achievable rate at $0$ dB receive SNR. As shown in Fig. \ref{fig:ach_rate_beam_set_size}, when the RIS sub-codebook size increases, the achievable rate performance of our proposed vision-aided RIS beam selection approaches the performance of the exhaustive search. \textbf{When only $\bf 12$ out of $\bf 256$ beams are used in the RIS sub-codebook, the proposed approach can achieve $\bf 97.4\%$ achievable rate compared with the exhaustive search}. Note that the exhaustive search provides the \textbf{optimal achievable rate} under the RIS beam codebook constraint. Therefore, approaching this upper bound indicates that the proposed vision-aided RIS beam selection can efficiently reduce the beam training overhead with little negative effect on the achievable rate of the RIS-aided system.

	\section{Conclusion}\label{Conclusion}
	In this paper, we explored a novel direction of aiding the RIS beam selection with visual sensors (cameras) deployed at the RIS. We developed an ML framework and a neural network architecture to predict the promising RIS beams using the visual sensing information. To evaluate the developed solutions, simulations were conducted based on a high-fidelity synthetic dataset gathering co-existing wireless and visual data. When benchmarked against  two other NN architectures, the proposed NN showed a clear advantage in both learning stability and beam prediction performance. In particular, simulation results demonstrated that the developed ML framework can accurately predict the RIS candidate beam set with high accuracy and recall performance. Further, the results showed that the proposed vision-aided RIS beam selection can obtain a near-optimal achievable rate with significantly reduced beam training overhead. This highlights the potential of leveraging visual sensors to enhance RIS communication systems.
	

\end{document}